\documentclass[%
 aip,
 amsmath,amssymb,
 reprint,%
]{revtex4-1}

\usepackage{graphicx}
\usepackage{dcolumn}
\usepackage{bm}

\usepackage{tikz}

\usepackage[utf8]{inputenc}
\usepackage[T1]{fontenc}
\usepackage{mathptmx}
\usepackage{xcolor}
\usepackage{amsmath, amssymb}
\usepackage{mathtools}
\usepackage{etoolbox}
\usepackage{qcircuit}
\usetikzlibrary{arrows, positioning}

\let\oldAA\AA
\renewcommand{\AA}{\text{\normalfont\oldAA}}

\makeatletter
\def\@email#1#2{%
 \endgroup
 \patchcmd{\titleblock@produce}
  {\frontmatter@RRAPformat}
  {\frontmatter@RRAPformat{\produce@RRAP{*#1\href{mailto:#2}{#2}}}\frontmatter@RRAPformat}
  {}{}
}%
\makeatother
\begin{document}

\preprint{AIP/123-QED}

\title{Fast-forwarding molecular ground state preparation with optimal control on analog quantum simulators}
\author{Davide Castaldo}
\affiliation{Università degli studi di Padova, Dipartimento di Scienze Chimiche, Via Marzolo 1 - 35131 Padova (Italy)}
\email{davide.castaldo@studenti.unipd.it}
\author{Marta Rosa}%
\affiliation{Università degli studi di Padova, Dipartimento di Scienze Chimiche, Via Marzolo 1 - 35131 Padova (Italy)}
\author{Stefano Corni}
\affiliation{Università degli studi di Padova, Dipartimento di Scienze Chimiche, Via Marzolo 1 - 35131 Padova (Italy)}
\affiliation
{Padua Quantum Technologies Research Center, Università di Padova}
\affiliation
{Istituto Nanoscienze—CNR, via Campi 213/A, 41125 Modena (Italy)}


\begin{abstract}

   We show that optimal control of the electron dynamics is able to prepare molecular ground states, within chemical accuracy, with evolution times approaching the bounds imposed by quantum mechanics.
  We propose a specific parameterization of the molecular evolution only in terms of interaction already present in the molecular Hamiltonian. Thus, the proposed method solely utilizes quantum simulation routines, retaining their favourable scalings. 
  Due to the intimate relationships between variational quantum algorithms and optimal control we compare, when possible, our results with state-of-the-art methods in literature. We found that the number of parameters needed to reach chemical accuracy and algorithmic scaling are in line with compact adaptive strategies to build variational ansatze.
  The algorithm, which is also suitable for quantum simulators, is implemented emulating a digital quantum processor (up to 16 qubits) and tested on different molecules and geometries spanning different degrees of electron correlation.
\end{abstract}

\maketitle

\section{\label{sec:level1} Introduction}

At the heart of the second quantum revolution are two main characters: those working to counteract the harmful effects of quantum noise and those seeking the most efficient strategies to gain practical advantage from new quantum devices as soon as possible. Both face this challenge because properly harnessing physics at the nanoscale would enable the leap forward envisaged by the use of quantum computers\cite{deutsch2020harnessing, feynman:1999, deutsch:1985}. First evidences of a quantum advantage\cite{arute2019quantum, zhong2020quantum, madsen2022quantum}, albeit with some \textit{caveat}\cite{zhou2020limits}, are now coming to light.

Within this framework, the precise manipulation of quantum matter is central to the emergence of new ideas leveraging non-classical properties. Coherent control, before being the basis of quantum information processing techniques \cite{werninghaus2021leakage, montangero2007robust}, has been proven a pivotal tool for the exploration of exotic states of matter enabling the preparation of ultracold atoms \cite{sklarz2002loading, koch2004stabilization} or the investigation of ultrafast electron dynamics \cite{klamroth2006optimal, rosa2019quantum}. Accomplishing these tasks means to push the boundaries of engineering into the realm of quantum physics, which has led to the rapid development, both in terms of techniques \cite{doria2011optimal, zhu1998rapidly} and applications \cite{brif:2010}, of Quantum Optimal Control Theory (QOCT) \cite{werschnik2007quantum}. In this context the physical knowledge of the system and the ability of simulating its evolution in presence of an external control are exploited to enhance a desired response by specifically tailoring a tunable perturbation.

Here we propose an optimal control approach to find the ground state of a molecular Hamiltonian: the real time evolution in presence of an external perturbation is handled by a quantum device, while the optimisation of the perturbation is carried out on a classical hardware to minimize the energy of the system. In this manner the computational burden of the simulation is addressed by the quantum device for which the solution of this task is expected to be one of the first applications with a significant advantage over classical hardwares \cite{lloyd1996universal, georgescu2014quantum}. The implementation of this routine does not require the universality of the quantum platform and is applicable to both an analog simulator and digital quantum computers. Our work will be focused on the problem of determining the ground state energy of a molecular system and the implementation that we will show directly relates to digital quantum simulators whose technology, to date, is more mature than the one of analog quantum simulators for chemistry \cite{arguello2019analogue, arguello2020quantum}. Nevertheless, given the potential of this latter alternative methodology, we will comment the possible development of our algorithm on an analog platform. 

Similar optimal protocols have already been applied to the case of laser cooling in a fully classical implementation \cite{rahmani2013cooling}. Differently from the standard laser cooling procedure (which is mostly related to the vibrational and rotational degrees of freedom \cite{bartana2001laser, bartana1997laser}), here we do not drive the evolution towards states that are prone to relax towards the target state but, instead, point directly to the target state minimizing the energy of a closed quantum system. For this reason we will refer to the proposed method as quantum simulated cooling since the ultimate goal of the algorithm is to find the optimal perturbation that realizes a trajectory (among those that can be realised by the adopted time-dependent Hamiltonian) driving the system from an initial guess state (higher in energy) to its ground state.

The algorithm we are proposing is based on the idea of solving an optimal control problem using a quantum system as a co-processor; is therefore inscribed in a research line where other problems have been addressed similarly. In particular, Li \textit{et al.} have considered a state preparation problem on a NMR quantum processor \cite{li2017hybrid}; Judson and Rabitz \cite{judson1992teaching} focused on the issue of optimal population transfer in ultrafast spectroscopy exploiting a closed-loop feedback control strategy. They proposed to shape the impinging light pulse on the basis of the molecular response until the evolution reaches optimally the desired state. Recently this idea has been extended, using a quantum computer, to those cases for which it is not possible to build such an experimental apparatus \cite{castaldo2021quantum}.  

In addition to these works, the relation between optimal control and variational hybrid algorithms is deep and has been discussed in Ref. \cite{magann2021pulses}. Among the plethora of ansatze proposed in literature to solve quantum chemistry problems, the works of Wecker \textit{et al.} \cite{wecker2015progress} and Choquette \textit{et al.} \cite{choquette2021quantum} are closely related to this work. The former has proposed, as ansatz for the Variational Quantum Eigensolver (VQE), a parametrized quantum circuit of the same structure of the system Hamiltonian allowing to restrict the variational search within a symmetry-preserving subspace. This eases the optimization that is challenging when occuring in the entire qubits register Hilbert space. The work of Ref. \cite{choquette2021quantum} moves from this point to include an additional term in the variational circuit accounting for temporary drifts in subspaces where the symmetries of the system are not conserved. Even though the variational circuit is not meant to realize a real time evolution of the system, the additional term that is included is thought as an external control highlighting, once again, the close link between optimal control and variational hybrid algorithms.

Finally, we also mention the work of Meitei \textit{et al.}\cite{meitei2020gate} that, very recently, has explored the possibility of rephrasing a VQE approach to optimize not a unitary generated by a parametrized quantum circuit (i.e., the typical approach) but rather to shape the state preparation unitary applying an optimal control protocol directly on the hardware Hamiltonian. Following up this work, many efforts\cite{meirom2022pansatz} have been done to reduce the duration of the pulses during the control solution either including the pulse length into the variational optimization\cite{egger2023pulse} or allowing leakages of the computer wavefunction outside the standard computational space\cite{asthana2022minimizing}. This work aims to contribute to these research lines considering an analog quantum simulator specifically devised for the molecular Hamiltonian.

This paper is organized as follows: in Sec. \ref{quantum_simulated_cooling} we present the general structure of the method, in particular, the main steps that must be followed when applying this procedure to any system are identified. In this regard, in section Sec. \ref{classical_optimization} we describe our choice for the energy optimization task on the classical hardware. Section \ref{molecular_application} presents applications to molecular systems describing in detail the control problem and the parameterization of the control operators (Sec.\ref{parameterization_sec}). Results and technical details about the implementation are provided in Sec. \ref{results}. We first show how the optimal control procedure is able to find the fundamental state while maintaining chemical accuracy. Then we focus on the effects that the length of the dynamics can have on optimization. We find that the dynamics has an optimal length in terms of convergence and result found, we compare these times of the dynamics to quantum speed limit estimates (i.e. minimum times to accomplish evolution according to quantum mechanics) for processes driven by time-dependent Hamiltonians, and, as reported in Ref.\cite{caneva2009optimal}, we find that theoretical bounds for time-dependent processes provide quite loose estimates compared to numerical results. 
Moreover, we study the convergence of the problem by keeping the evolution duration fixed and increasing the number of control parameters. Finally, section\,\ref{cost-section} is devoted to a semi-empirical estimate of the computational cost of this method obtained applying this method to hydrogen chains of different lengths. To conclude, we summarize the results obtained and discuss potential future extensions of this work. 

\section{Quantum simulated cooling}\label{quantum_simulated_cooling}

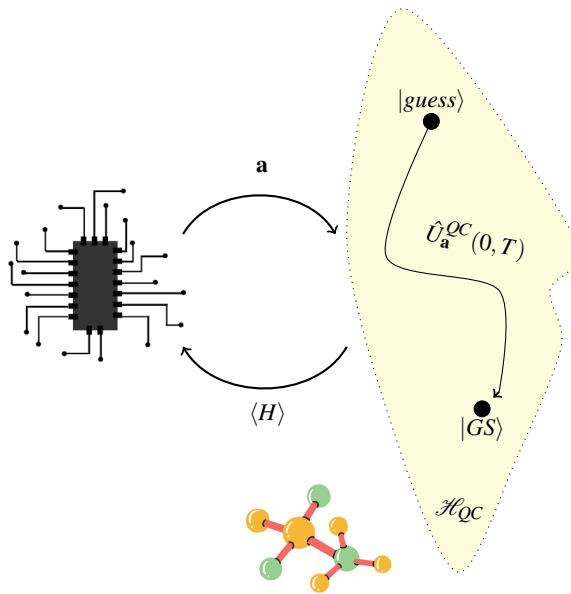
\begin{figure*}[htbp]

\centering

\begin{tikzpicture}[scale=1.5]

\definecolor{giallino_panna}{HTML}{FCE825}

\node (figure_1) at (-2,-0.75) {\includegraphics[scale = 0.3]{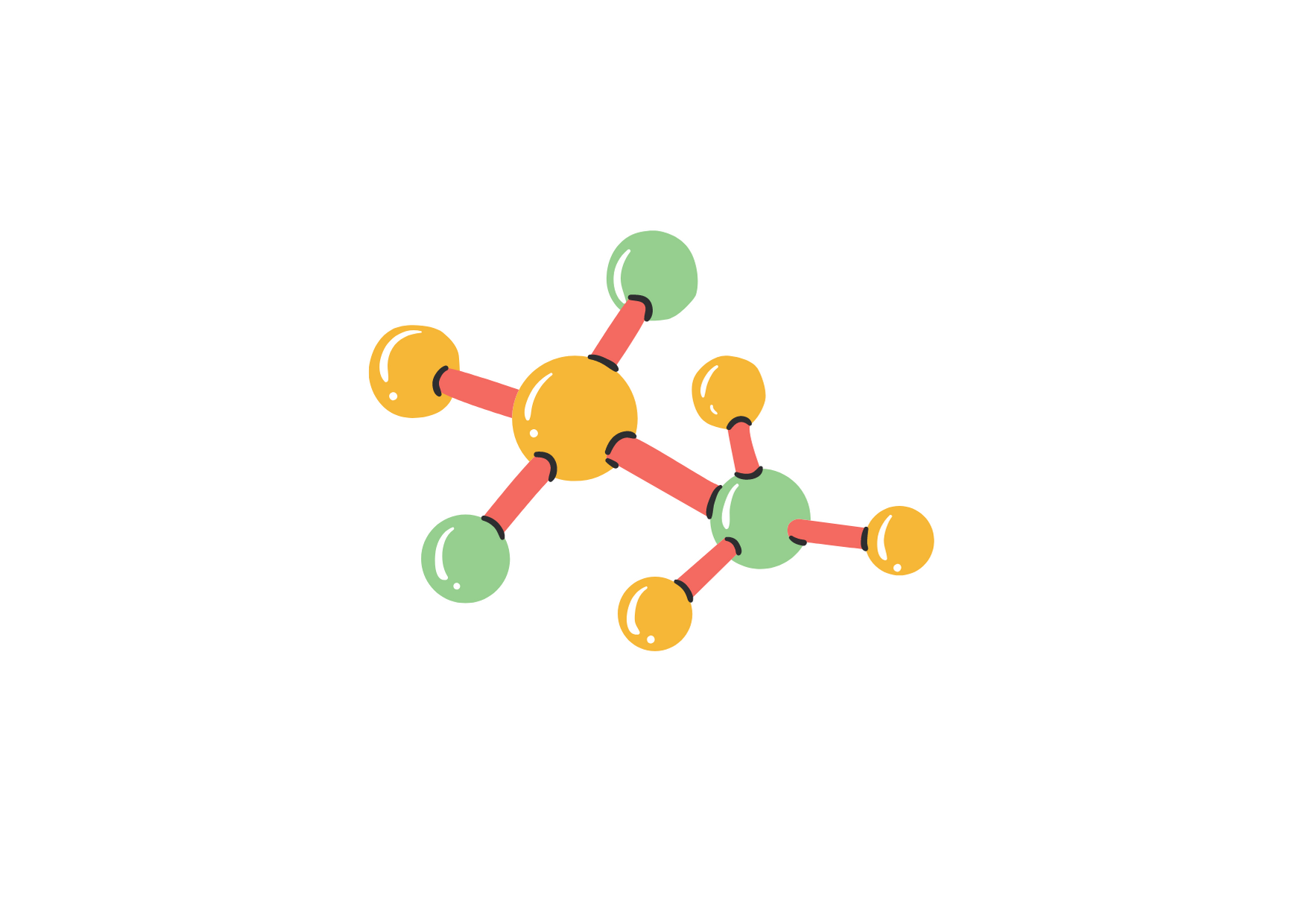}
};

\node (figure_2) at (-3.5,1.65)    {
    \includegraphics[scale = 1.1]{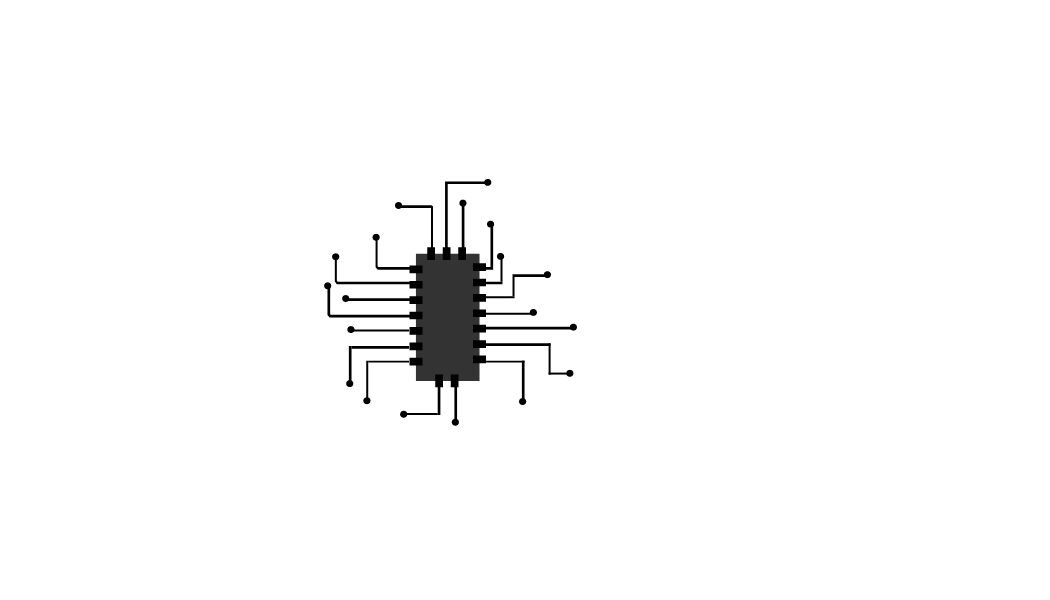}
};

 
 

 




\draw[rotate = 90, dotted,fill=giallino_panna!15]
plot [smooth cycle]
coordinates {(-1,.75) (0,1.25) (2,1.75) (4,1.25) (2,-.25) (1.5,-.05) (1,-.15) (0,.25)};


\filldraw [rotate = 90] (3,1) circle (2pt)
node[above] {$|guess\rangle$};

\filldraw [rotate = 90] (0.45,0.55) circle (2pt) node[below] {$|GS\rangle$};



\draw plot [rotate = 90, smooth] coordinates {(3,1) (1.8, 1.4) (1.5, 0.4) (0.6,0.4)};

\draw [->, rotate = 90, thick] {(0.6, 0.4) to (0.55, 0.45)};



\node [rotate = -10] (a) at (-0.6, 1.95) {$\hat{U}^{QC}_{\textbf{a}}(0, T)$};


\node (c) at (-0.75, -0.45) {$\mathcal{H}_{QC}$};


\draw [->, thick] {(-3.2, 2) to [out=60, in = 120] (-1.85, 2)} node at (-2.5, 2.6)  {$\textbf{a}$};

\draw [<-, thick] {(-3.2, 1) to [out=-60, in = -120] (-1.75, 1) } node at (-2.45, 0.4)  {$\langle H \rangle$};

\end{tikzpicture}
    \caption{Schematic diagram for the hybrid algorithm.  The evolution of the system wavefunction is performed on a quantum simulator, the Hamiltonian expection value $\langle H \rangle$ is measured feeding a classical optimization routine which outputs a new set of control parameters \textbf{a}. The control parameters shape the evolution specified by the time evolution operator $\hat{U}^{QC}_{\textbf{a}}(0, T)$ which starts from an higher energy initial guess state, $|guess\rangle$, and drives the evolution towards the exact ground state of the system, $|GS\rangle$. The loop ends when the energy is below a user-specified threshold.}
    \label{schema}
\end{figure*}

In this work we assume that a reference wavefunction $|\Psi_0\rangle$, approximating the target ground state $|\Psi_{GS}\rangle$ of the problem Hamiltonian $\hat{H}$, can be computed efficiently with a preliminar classical computation (e.g. such as the solution of a mean field effective Hamiltonian). We propose the use of a quantum processor to simulate the dynamics of the system Hamiltonian in presence of a time-dependent external control $\hat{V}(t)$. If the unitarity of the dynamics is preserved, $\hat{V}(t)$
can be adapted differently depending on the problem at hand, with the purpose of finding the ground state of the system. The perturbation can  either represent a real physical process (such as the presence of an external field coupled to some system's degree of freedom) or a process without experimental counterpart, not affecting the viability of the proposed computational method.  Once the evolution is computed, the quantum processor is used to evaluate the Hamiltonian expectation value for the system at the final time $t = T$, then the perturbation is iteratively shaped by a classical optimizer which aims to minimize the energy of the system (see Fig. \ref{schema}).

More formally, the quantum computer provides the evolution of the system due to a parametrized Hamiltonian $\hat{H}_{\textbf{a}}(t)$:

\begin{equation}
\label{main_eq}
    \hat{H}_{\textbf{a}}(t) = \hat{H} + \hat{V}_{\textbf{a}}(t)
\end{equation}
Where $\textbf{a}$ is the set of control parameters shaping the external perturbation.

In order to compute the evolution, the simulated Hamiltonian is mapped onto a N-qubits quantum register:

\begin{equation}
    \hat{H}^{QC}_{\textbf{a}}(t) = \sum_{j} \gamma^{\textbf{a}}_j(t) \hat{P}_j
\end{equation}
Here $j$ is an index running on different Pauli strings $\hat{P}_j \in \{ \sigma_x, \sigma_y, \sigma_z, \mathbb{I} \}^{\otimes N}$ that are operators acting non trivially on $k$ different qubits (in order to ensure the simulation routine efficiency \cite{lloyd1996universal}). The coefficients $\gamma^{\textbf{a}}_j(t)$ include both the system Hamiltonian and time dependent perturbation matrix elements, whose explicit form (as for the Pauli strings) depends on the adopted mapping.

Notably, we have not made any observation regarding the quantum simulation paradigm to adopt,  as 
the application of this methodology is not limited to universal quantum computers but can be applied with any suitable quantum simulator. In the following we will make explicit reference to the implementation on a digital quantum computer, bearing in mind that other routes are available which may be more or less convenient depending on the system under consideration.


Several methods have been developed to implement the evolution of a time dependent Hamiltonian \cite{wiebe2010higher, berry2015simulating, low2017optimal} here we compute the approximate time evolution operator $\hat{U}(0, T)$, discretizing the time axis with K slots of width $\Delta t = \frac{T}{K}$. Within each time slot we consider the Hamiltonian as time independent. The accuracy of this procedure depends on the precision with which the Hamiltonian is simulated within each time slot, the time step used and the complexity of the perturbation \cite{wiebe2011simulating}. 



So far we have presented the general framework of the method and discussed the role played by the quantum device. In the following section we discuss the complementary step of the procedure: the classical update of the control parameters.

\subsection{Classical optimization of the energy functional}\label{classical_optimization}


In the framework of variational hybrid algorithms the choice of the classical optimization routine is a crucial step \cite{mcardle2020quantum} not only on its own but also with respect to the quantum resource requirements. Indeed, as a first approximation we can estimate the cost of running the optimal control problem to find the ground state as $C=\mathcal{O}(\frac{K M G}{\epsilon^2})$, where $K$ is the number of iterations needed to achieve convergence of the result, $M$ is the number of circuits one needs to execute to do one step of the optimization and $G$ is the gate count of each circuit. The term $\epsilon^{-2}$ comes from the finite-shot sampling noise. We will discuss the computational scaling of the control protocol proposed in this work in Sec.\ref{cost-section}, here we want to discuss only the possible choices of the optimizer that impact primarily the quantity $K$.

For our purposes the cost function that we want to minimize is given by:

\begin{equation}
    J[\textbf{a}] = \langle \Psi_{\textbf{a}}(T) | \hat{H} | \Psi_{\textbf{a}}(T) \rangle
\end{equation}
Where $|\Psi_{\textbf{a}}(T) \rangle$ is the wavefunction of our system of interest at time T after the application of the parameterized evolution $\hat{U}_{\textbf{a}}(0, T)$.

If we consider current quantum processors, it is reasonable to prefer gradient-free optimizers as the calculation of functional gradients numerically or exactly\cite{schuld2019evaluating,mari2021estimating,mitarai2018quantum} would imply adding further noise sources. 
On the other hand, trying to ensure scalability of the algorithm requires that the optimization task to be accomplished into a reasonable number of iterations to avoid the circuit number executions to grow too rapidly. To this extent the most natural option would be using a gradient-based optimization. Indeed, it is known that the convergence rate for many optimization problems is higher for these latter kind of algorithms than for gradient-free optimizers\cite{machnes2018tunable}. As described in Sec.\ref{computational_details} here we adopted the L-BFGS\cite{wright2006numerical} optimizer.

Before concluding this section it is important to remark that the close relationship between optimal control, variational algorithms and supervised learning may lead to improvements of our implementation that in turn can reduce the overall scaling. Particularly, very recently natural gradient based methods\cite{stokes2020quantum, fitzek2023optimizing} have been developed in the context of VQAs showing promising results in terms convergence rate and avoidance of barren plateaux. Further, other strategies implementing reinforcement learning techniques have shown great improvements w.r.t. gradient based methods that could possibly lead to further speedups to the proposed algorithm\cite{dalgaard2020global, niu2019universal, wright2023fast}. We will seek to explore these aspects in future contributions.

\section{Molecular ground state energies}
\label{molecular_application}

In this section we provide a detailed description of the algorithm sketched above applied to the case of molecular systems. Therefore, analogue to the previous section, we identify with the problem hamiltonian $H$ the molecular hamiltonian $H_{mol}$:

\begin{equation}
\label{molecular_hamiltonian}
    \hat{H}_{mol} = \sum_{p,q} h_{pq} a^{\dagger}_{p}a_q + \frac{1}{2}\sum_{p,q,r,s} g_{pqrs} a^{\dagger}_{p}a^{\dagger}_{r}a_{s}a_{q}
\end{equation}
Where $h_{pq}$ are the one-electron integrals containing the kinetic energy and the electron-nuclei repulsion terms and $g_{pqrs}$ are the two-electron repulsion integrals. 

To compose the parametrized Hamiltonian $\hat{H}_{\textbf{a}}(t)$ we have to specify a form for the perturbation operator and a proper parametrization to implement the optimization routine. As already mentioned in Sec. \ref{quantum_simulated_cooling} different options are viable, here we considered a time-dependent modification of the Hamiltonian expressed in terms of five ingredients: (i) an effective electron mass $m_e(t)$, (ii) effective nuclear charges $\tilde{Z}_i(t)$, (iii) screened electron-electron interactions $\tilde{\epsilon}(t)$, (iv) a time-dependent effective scalar mean field term $b_0(t)$ and, finally, (v) an overall scalar prefactor $a_0(t)$. These lead to the following expression for $\hat{V}_{\textbf{a}}(t)$:

\begin{equation}
\label{perturbation}
    \hat{V}_{\textbf{a}}(t) = a_0(t) \Big [ \sum_{p,q} \tilde{h}^{\textbf{a}}_{pq} (t) a^{\dagger}_{p}a_q + \frac{1}{2}\sum_{p,q,r,s} \tilde{g}^{\textbf{a}}_{pqrs} (t) a^{\dagger}_{p}a^{\dagger}_{r}a_{s}a_{q} \Big ]
\end{equation}

where $\tilde{h}^{\textbf{a}}_{pq} (t)$ and $\tilde{g}^{\textbf{a}}_{pqrs} (t)$ are given by:

\begin{widetext}
\begin{equation}
 \tilde{h}^{\textbf{a}}_{pq} (t) = 
    \begin{cases}
   \frac{1}{2 m_e(t)}\int\phi^*_p(\textbf{x})\nabla^2\phi_q(\textbf{x})d\textbf{x} - \int\phi^*_p(\textbf{x})\sum_i \frac{\tilde{Z}_i (t)}{r_i}\phi_q(\textbf{x})d\textbf{x} & p \neq q \\
      (b_0(t) + \frac{1}{2 m_e(t)})\int\phi^*_p(\textbf{x})\nabla^2\phi_q(\textbf{x})d\textbf{x} - b_0(t)\sum_i(Z_i + \frac{\tilde{Z}_i(t)}{b_0(t)})\int\phi^*_p(\textbf{x}) \frac{1}{r_i}\phi_q(\textbf{x})d\textbf{x} & p = q \\
    \end{cases}
\end{equation}
\end{widetext}

and

\begin{equation}
 \tilde{g}^{\textbf{a}}_{pqrs} (t) = 
    \begin{cases}
   (\frac{1}{\tilde{\epsilon}(t)} +  b_0(t))   g_{pqrs}  & p = r, q = s \\
    (\frac{1}{\tilde{\epsilon}(t)} - b_0(t)) g_{pqrs} & p = s, q = r \\
     \frac{1}{\tilde{\epsilon}(t)} g_{pqrs} & \text{else}  \\
    \end{cases}
\end{equation}

The choice of this parameterization is guided by three factors: (i) linking the perturbation explicitly to physical quantities that are accessible in the presence of an analog simulator; (ii) minimizing the number of parameters to be optimized to reduce the computational cost of optimization; and (iii) ensuring the necessary expressivity for the perturbation to effectively generate dynamics that lead to the target state. Particularly, the last principle motivated the choice of a differential treatment for each atom in the molecule: in fact, it may allow a more subtle discrimination between spatial orbitals (that already experience different nuclear charge due to $a_0$ and $b_0$ factors) enhancing effects due to the distinct atoms electronegativity. The extent to which this feature impacts the optimization of the wavefunction is an intriguing question in itself, as it may provide additional physical insight into the solution of the control problem. We aim to delve deeper into investigating this particular feature in a follow-up study.

 Concerning the trainability of the proposed parameterization,  it is important to mention the problem of barren plateaus which is of specific relevance to hybrid variational algorithms \cite{mcclean2018barren, cerezo2021cost}. It has been shown that the landscape parameters' shape is strongly affected by the exponentially big dimension of the quantum processor's Hilbert space. 
More precisely, as we consider larger systems (i.e. a greater number of qubits), the average value of the gradient objective function tends to zero and more and more states embody this typical value. Thus, if we do not leverage the physical intuition coming from the model Hamiltonian of interest (e.g. exploit symmetry constraint), the control parameters landscape becomes flat over a larger portion of the Hilbert space that we explore during our optimization procedure. We will discuss this aspect in more detail in Sec.\,\ref{qsl-sec}.

\subsubsection{Parametrization of the control Hamiltonian}\label{parameterization_sec}

Now we turn our attention to the parametrization of the external control. Within the context of quantum optimal control various shapes for the control fields have been proposed ranging from superposition of Gaussian pulses \cite{machnes2018tunable} and Fourier-based parametrizations (such as the CRAB and DCRAB methods \cite{caneva2011chopped, rach2015dressing}) to a point-wise definition of the temporal profile as in the case of the GRAPE algorithm \cite{khaneja2005optimal}. In this work, we have opted to represent the control functions using the latter option. Previous studies that applied similar parameterizations to optimal gate synthesis have demonstrated that this approach is not always preferable compared to the analytic parameterization\cite{willsch2017gate}. Indeed, analytic controls may allow a faster computation of the gradients and have been shown to be less prone to introducing unwanted high-frequency components leading to leakage errors. However, when transitioning to experimental setups where discretization of control pulses becomes inevitable, these disadvantages fall short\cite{niu2019universal}. Since we wanted to focus on the development of an algorithm as more oriented to analog simulators as possible we focused on this approach.

With these choice of the parameterization we get an explicit scaling of the number of parameters w.r.t. the time steps of the evolution and system size which is $\mathcal{O}(M T)$. Where $M$ is the number of nuclei and $T$ is the number of controlled steps of the discretized evolution. In Sec.\ref{cost-section} we will provide semi-empirical estimates of the scaling w.r.t. the number of spin-orbitals showing results on hydrogen chains of different length.

It is worth noticing that the energy expectation value measured at the end of the perturbation, as mentioned in Sec. \ref{quantum_simulated_cooling}, is related to the system Hamiltonian $\hat{H}_{mol}$. Hence, no boundary conditions on the perturbation that impose the driving Hamiltonian to coincide with the system Hamiltonian at the end of the evolution are needed. Finally, in contrast with usual optimal control protocols applied to laboratory experiments, unless it is useful for the optimization, the control parameters are allowed to take arbitrary values without maximum (or minimum) thresholds or penalty. 

\section{Results}
\label{results}

\subsubsection{Computational details}\label{computational_details}

In this section we provide the computational details for the implementation of all our numerical results shown in Sec. \ref{oc-energies}-\ref{cost-section}. All the calculations were performed with a Python code using a locally modified version of the PennyLane library\cite{bergholm2018pennylane} to construct a representation of all the operators in the computational basis. The evolution and the optimization of the wavefunction were carried out using JAX\cite{jax2018github} and the JaxOpt\cite{jaxopt_implicit_diff} library in order to exploit automatic differentiation and fast evaluation of the quantum dynamics with just-in-time compilation of the code. To ease the computational burden of exactly simulating the quantum dynamics into the qubit space we adopted symmetry reduction of the operator representation to taper-off redundant qubits as shown in Ref.\cite{setia2020reducing} and implemented in \cite{bergholm2018pennylane}. The code is available open-source at \cite{oc_code}. Regarding the specifics of the optimization we used the L-BFGS algorithm as implemented in JaxOpt with default settings. All the calculations were performed either with a maximum number of iterations (N$_{iter}$) or an energy error of 1 mHa as termination condition. 

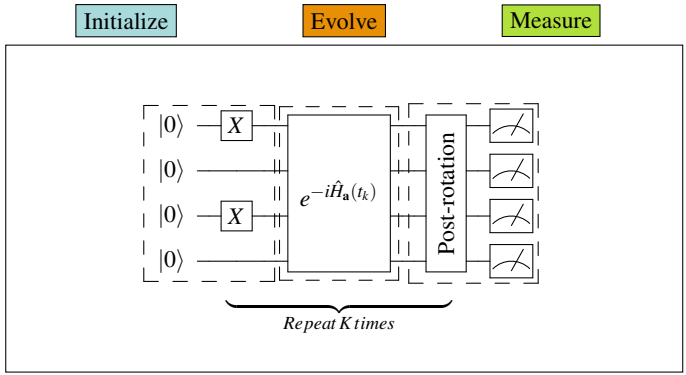
\begin{figure}[h!]
\centering
\begin{tikzpicture}

\definecolor{arancio}{HTML}{E89005}
\definecolor{azzurro}{HTML}{457B9D}
\definecolor{azzurrino}{HTML}{A8DADC}
\definecolor{tartaruga}{HTML}{B1E03C}

    \node[draw, fill=azzurrino] at (-8.9,2.5) {Initialize};

    \node[draw, fill=arancio] at (-6,2.5) {Evolve};

    \node[draw, fill=tartaruga] at (-3.25,2.5) {Measure};
    
    
    
    
    
    

  \node (Quantum Routine) at (-6, 0) [draw, minimum height = 4.35cm, minimum width = 9cm] { \Qcircuit @C=.5em @R=.5em {  \lstick{|0\rangle} & \qw & \gate{X}
 &\qw & \qw & \multigate{3}{e^{-i \hat{H}_{\textbf{a}}(t_k)}} & \qw &\qw & \multigate{3}{\rotatebox{90}{Post-rotation}} &\qw & \meter \\
\lstick{|0\rangle} & \qw & \qw &\qw & \qw & \ghost{e^{-i \hat{H}_{\textbf{a}}(t_k)}} & \qw &\qw & \ghost{\rotatebox{90}{Post-rotation}} & \qw &\meter\\
 \lstick{|0\rangle} & \qw & \gate{X} &\qw & \qw & \ghost{e^{-i \hat{H}_{\textbf{a}}(t_k)}} & \qw &\qw & \ghost{\rotatebox{90}{Post-rotation}} &\qw & \meter \\
\lstick{|0\rangle} & \qw & \qw &\qw &\qw & \ghost{e^{-i \hat{H}_{\textbf{a}}(t_k)}} & \qw &\qw & \ghost{\rotatebox{90}{Post-rotation}} & \qw &\meter
& {\gategroup{2}{2}{3}{2}{5.5em}{--}} \\
&  {\gategroup{1}{6}{4}{6}{.7em}{--}}\\
&  {\gategroup{1}{8}{4}{11}{.45em}{--}}\\
& & & & & \relax{\underbrace{\hspace{3cm}}_{Repeat \, K \, times}} }};

\end{tikzpicture}
    \caption{Example circuit needed for the implementation on a digital quantum processor. Quantum computer's initial state is given by all the qubits being in the state $|0\rangle$; as an example we reported the initialization circuit for the $H_2$ molecule in the minimal basis. The second step provides the evolution of the molecular wavefunction, due to the time-dependent external control the exponentiation of the Hamiltonian is repeated for K different steps of the propagation. Here we mantained the circuit as more general as possible (i) to resemble the numerical exponentiation that we used in the computational protocol and (ii) to highlight that one can choose the more suitable simulation routine at hand. Finally the circuit is repeated several times to evaluate the hamiltonian expectation value.}
    \label{quantum_simulation_circuits}
\end{figure}

The initial state for all our simulations is the Hartree-Fock wavefunction using the STO-3G basis set in all cases. The initial guess parameters were drawn from a uniform distribution between 0 and 1. The parameters $a_0$ and $b_0$ were initialized according to a linear schedule (see Appendix\,A) as is done by adiabatic state preparation protocols of varying lengths (depending on the time step of the simulation), ranging from 0.2 to 7.5 atomic units (a.u.).

In Fig. \ref{quantum_simulation_circuits} we report the general structure of the quantum circuits adopted in all the calculations. The qubits are initialized in the $|0\rangle$ state, the mapping between the qubits and the molecular spin-orbitals is accomplished according to the Jordan-Wigner method \cite{jordan:1928} (e.g. each occupied spin-orbital is represented by a qubit in the state $|1\rangle$). As previously mentioned in Sec. \ref{quantum_simulated_cooling}, the digital quantum simulation is performed numerically exponentiating the time-dependent hamiltonian at each time step in the computational basis spanned by the qubit register. The time step used varies between $\Delta t $ = 0.00125 a.u. and $\Delta t $ = 0.05 a.u. for all the simulation reported in this work. The stability of the numerical integration is assessed on the basis of previous works\cite{castaldo2021quantum}.

\subsection{Molecular ground state energies}\label{oc-energies}

In this section we report numerical examples of the control protocol applied to three different molecular systems: the $H_4$ molecule with atoms arranged in a square lattice, $H_6$ in a linear configuration and lithium hydride (LiH).

The selection of these systems was made to specifically assess the algorithm's performance on systems that, despite their small size, are well-established benchmarks for quantum chemistry methods. Notably, hydrogen chains, although experimentally unstable\cite{stella2011strong}, have been extensively characterized being the simplest systems revealing strong electronic correlation phenomena. We opted for two distances, r=1 $\AA$ and r=2 $\AA$, since the former is both in proximity of the observed metal-to-insulator phase transition point expressed in longer analogues of the same series\cite{motta2020ground} and results from other quantum variational methods are available for comparison. The geometry associated with r=2 $\AA$, which is farther from the equilibrium bond distance, allows us to put our method to the test in a regime approaching dissociation. On the other hand, the H$_4$ molecule in squared configuration is another prototypical system used to study multireference effects. At this geometry HOMO and LUMO orbitals become degenerate giving a diradical character to the electronic system\,\cite{sand2013parametric}.
Farther from equilibrium, the second geometry we have considered, on top of these effects we add a fourfold bond dissociation process that requires multiply excited configurations to be described. Finally, we have also considered lithium hydride as to test our parameterization (explicitly involving nuclear charges) with an heteroatomic system.

\begin{figure*}[htbp!]

\centering
    \includegraphics[width = \textwidth]{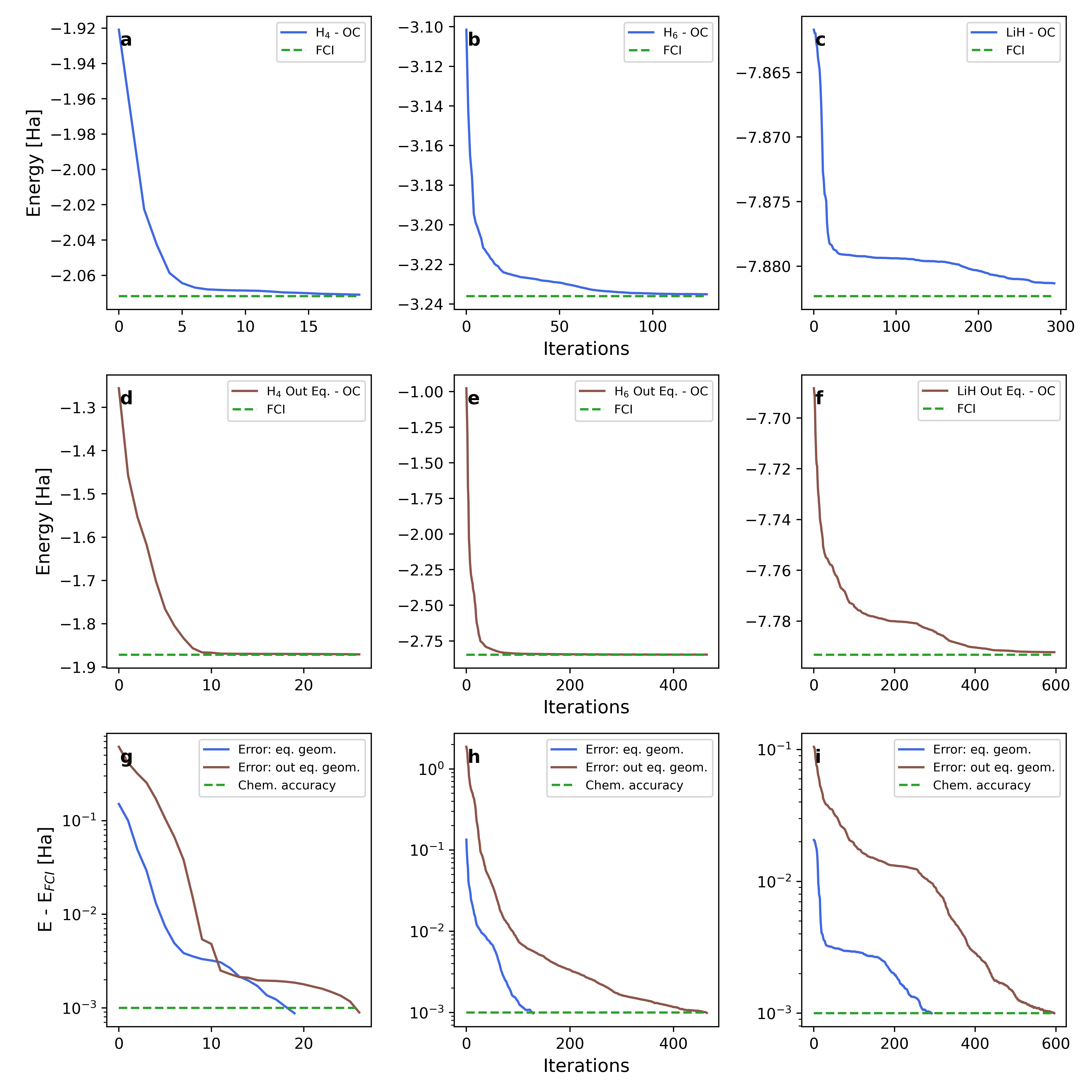}

    \caption{Optimal control for molecular ground state preparation. a-c) Energy (blue solid line) vs. iterations for squared H$_4$ at r = 1.2, $\AA$; linear H$_6$ chain at r = 1 $\AA$ and LiH at r = 1.6 $\AA$. d-f) Energy (brown solid line) vs. iterations for squared H$_4$ at r = 2.4 $\AA$; linear H$_6$ chain at r = 2 $\AA$ and LiH at r = 3.2 $\AA$. Error (dashed-dotted line) w.r.t. FCI Energy vs. iterations. Same color code as in panels above; green dashed lines represent either the FCI energy value or a threshold for chemical accuracy posed at 1mHa.}
    \label{oc_energies}
\end{figure*}

In Fig.\ref{oc_energies} we show the result of the control problem solution for the systems described above. These calculations demonstrate that the control problem can be effectively resolved, achieving error energies below 1 mHa ($< 0.67 \,\text{kcal/mol}$), even in regimes characterized by strong correlation. From a dynamical perspective, this implies our capability to identify a pathway leading from the HF state to the exact ground state, even when these states are significantly separated within the Hilbert space. However, it is worth noting that commencing the optimization process from a state further away in the Hilbert space tends to prolong the optimization, despite the protocol's capacity to reach a chemically accurate state. In this regard, in Section \ref{cost-section}, we have examined the convergence rate's scaling with respect to the system size to assess the computational scalability of this approach.

Before discussing these aspects we focus on the effect of varying the duration of the controlled dynamics.

\subsection{Ground state preparation at the quantum speed limit}\label{qsl-sec}

Quantum speed limits define the minimum time needed for a quantum system to transition between states. In quantum technologies, they have been extensively studied\cite{deffner2017quantum} as being able to engineer transformations achieving this boundaries directly impacts the efficiency and capabilities of quantum devices.

Several theoretical bounds have been developed to quantify this times for different kind of processes\cite{margolus1998maximum, chau2010tight, ness2022quantum}; here, following Ref.\,\cite{caneva2009optimal} we estimated the quantum speed limit $T_{QSL}$ (see Tab.\,\ref{qsl_tab}) for the optimally controlled trajectory as:
\begin{equation}
    T_{QSL} \leq \frac{\pi}{ 2 }\frac{T}{\int_0^T \sqrt{\langle \psi (t) | [\hat{H}(t) - E(t)]^2 | \psi(t) \rangle} dt} 
    \label{qsl_bound}
\end{equation}
Where $E(t) = \langle \psi(t) | H | \psi(t) \rangle$.

This quantity estimates the quantum speed limit as the mean energy spread along the computed trajectory and represent an extension of the Battcharayya bound\cite{bhattacharyya1983quantum} to time-dependent hamiltonians. Please notice that T$_{QSL}$ depends on the control parameters \textbf{a} as $|\psi(t)\rangle = U_{\textbf{a}}(0, t)|\psi(0)\rangle$. 

As reported in Ref.\,\cite{caneva2009optimal}, a rigorous definition of a quantum speed limit for time-dependent processes is elusive and discrepancies with numerical experiments reflect this aspect. The authors find for the problem of optimal population transfer along a spin chain that Eq.\,\ref{qsl_bound} overestimates (on average) by a factor of 3 the numerical results. These discrepancies have been shown with even tighter bounds as reported in Ref.\,\cite{lee2018dependence} for the case of a time-optimal SWAP gate. Here we found that the numerical estimate is sensibly lower than the theoretical one in all cases with edge cases such as H$_4$ (r = 1.2 $\AA$) where Eq.\,\ref{qsl_bound} overestimates the numerical result by two order of magnitudes. 

We can compare these findings with other similar works in literature. Particularly, the work of Matsuura et al.\cite{matsuura2020vanqver} proposes the coupling of an optimal control of the annealing schedule and a VQE with a UCCSD ansatz to improve molecular ground state preparation which they dub VanQver. We find similar times for the optimized evolution as they report. Particularly, for the rectangular H$_4$ molecule (notice that here we qualitatively compare the results as we have instead a perfectly squared geometry) they find T$_{VanQver} = 0.088 \, a.u.$, for the LiH molecule T$_{VanQver} = 0.14\, a.u.$. For comparison their reported times for Annealing State Preparation (ASP) are respectively $T_{ASP} = 9.5\, a.u.$ and $T_{ASP} = 11.5 \,a.u.$ for LiH and rectangular H$_4$ both close to the equilibrium bond length (i.e. to compare with the first and third column of Tab. \ref{qsl_tab}).

We can also compare the number of parameters needed by our optimal control procedure with the number of parameters generated by the UCCSD ansatz and compact adaptive strategies\cite{tang2021qubit,shkolnikov2023avoiding}. Particularly, as discussed in \cite{grimsley2019adaptive} for the LiH molecule, spin-adapted UCCSD requires 64 parameters in comparison adaptive ansatze built with a fermionic operator pool require around 10 parameters to reach chemical accuracy. In the same work the authors report, concerning the H$_6$ molecule, that the UCCSD circuit requires almost 70 parameters without being able to reach chemically accurate results as the bond distance increases (at about distances greater than 1.3 $\AA$) while the adaptive procedure requires at most the same number of parameters reaching chemical accuracy even in the dissociation limit. As we can see, the optimal control procedure requires a similar set of parameters as the adaptive approach when considering bond lengths close to the equilibrium geometry. However, in cases involving more stretched bonds, it becomes evident that the adaptive procedure shows greater efficiency.

       \begin{table*}
           \centering
            \caption{Estimated time according to the Bhattacharyya bounds\cite{bhattacharyya1983quantum, caneva2009optimal} (Eq. \ref{qsl_bound}) on time-dependent quantum evolutions ($T_{QSL}$), time required by the optimally controlled evolution to reach the ground state ($T_{OC}$) and number of parameters optimized in each trajectory (\# $ \theta_{QSL}$) for the molecules considered in this study. Time is expressed in atomic units.}
           \begin{tabular}{lc|c|c|c|c|c|c}
                & LiH @ 1.6 $\AA$ & H$_{4}$ @ 1.2  $\AA$ & H$_6$ @ 1 $\AA$ & LiH @ 3.2 $\AA$ & H$_{4}$ @ 2.4 $\AA$ & H$_6$ @ 2 $\AA$ \\
                \hline\hline
              T$_{QSL}$  & 7.91 & 2.39 & 6.27 & 10.84 & 7.43 & 39.11 \\\hline
              T$_{OC}$  & 0.25 & 0.01 & 0.5 & 0.75 & 0.5 & 0.75 \\\hline
             $\# \theta_{QSL}$ & 30 & 32 & 50 & 90 & 80 & 150 \\
           \end{tabular}

           \label{qsl_tab}
       \end{table*}

\begin{figure*}[htbp!]

\centering
    \includegraphics[scale = 0.75]{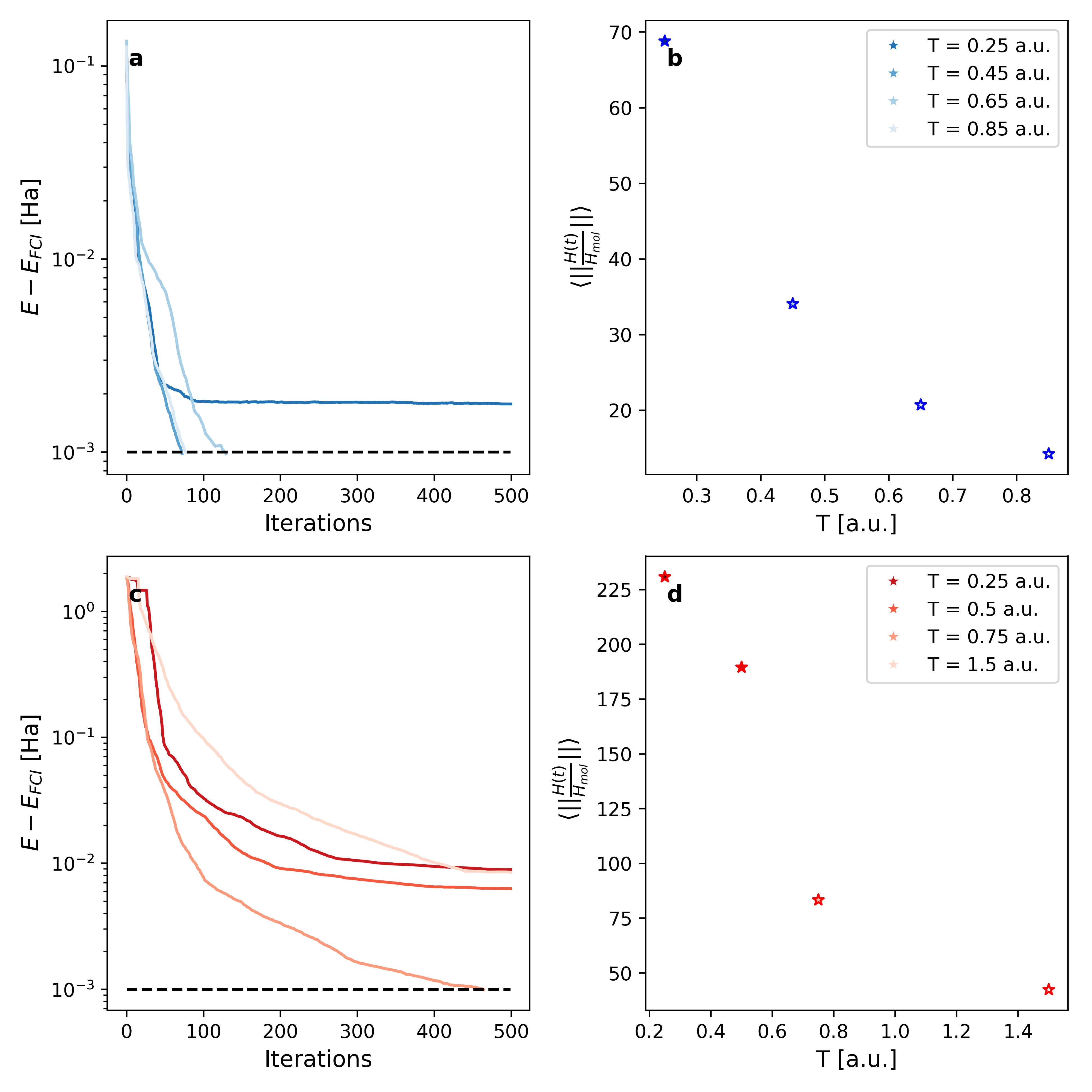}

    \caption{Ground state preparation at the quantum speed limit for the H$_6$ molecule. a, c) Optimal control for various duration of the dynamics. Blues refer to H$_6$ at r=1 $\AA$, reds refer to H$_6$ at r=2 $\AA$. b,d) Mean driving Hamiltonian norm vs. duration length for the same systems and color codes.}
    \label{qsl_fig}
\end{figure*}

\begin{figure*}[htbp!]
\centering
    \includegraphics[scale = 0.75]{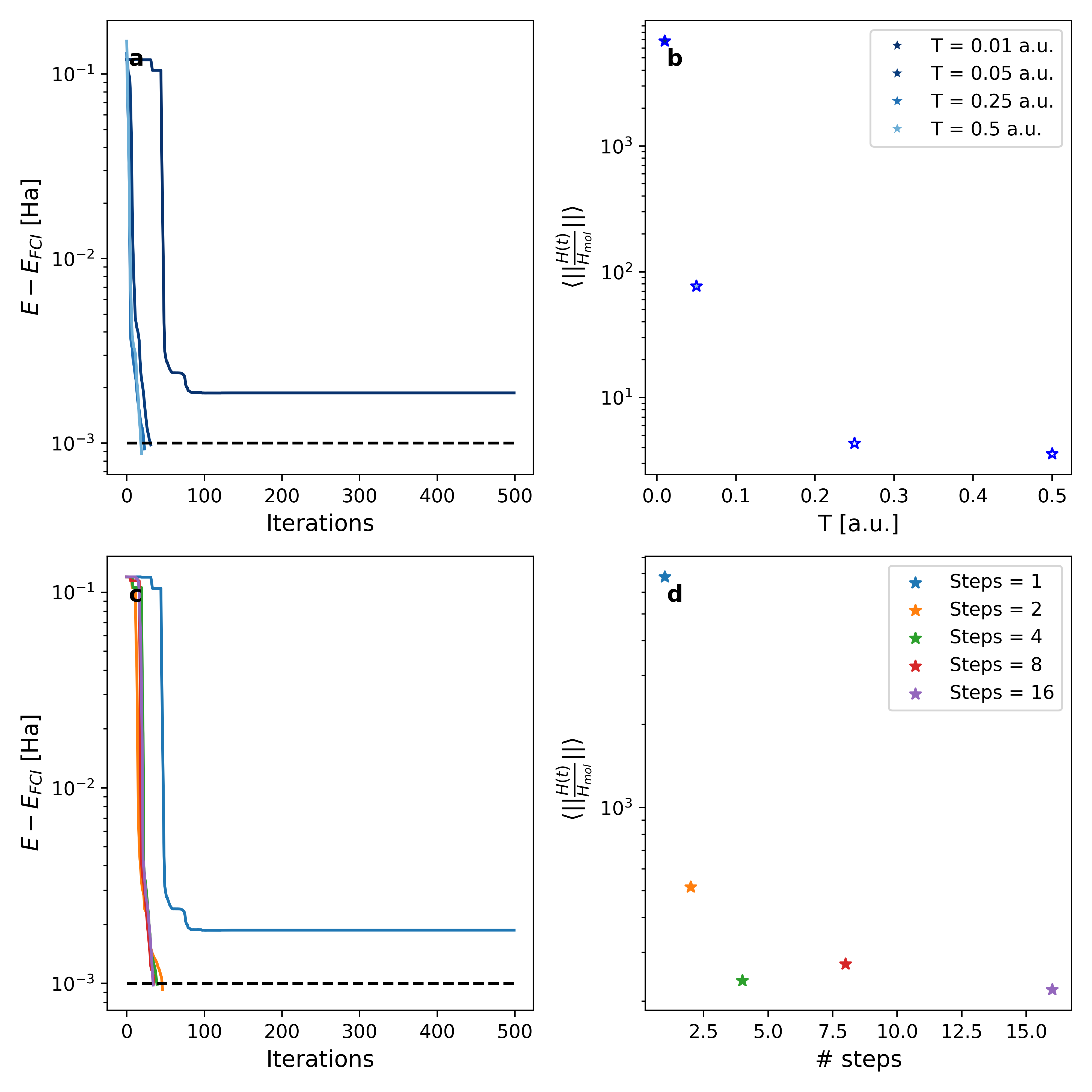}

    \caption{Effect of controllability on the time-energy uncertainty relationship. a) Optimal control for the H$_4$ molecule at r=1.2 $\AA$ for different quantum dynamics' lengths (darker blue shorter length). b) Mean driving Hamiltonian norm vs. duration length. c, d) Control problem at fixed duration length (T=0.01 a.u.) with finer time stepping (more controllable dynamics).}
    \label{controllability_qsl}
\end{figure*}

In addition to estimating the quantum speed limit through equation \ref{qsl_bound}, we analyzed the relationship between optimal control procedures for the systems presented in the previous section and the duration of the dynamics (see Fig. \ref{qsl_fig}). What is observed is a dual behavior: concerning dynamics shorter than a certain length (which we identify as the actual quantum speed limit), the optimal control procedure is unable to achieve chemical accuracy within 500 iterations. Furthermore, the optimizer amplifies the perturbation strength by increasing the average energy injected into the system, which we tried to estimate with the quantity $\langle \frac{||H(t)||}{||H_{mol}||} \rangle$  (see Fig.\ref{qsl_fig}-\ref{controllability_qsl}). As we can see, consistently with the time-energy uncertainty relationship, the average energy injected into the system decreases as the evolution length increases.

On the other hand, the optimized dynamics with a duration exceeding the quantum speed limit converges more slowly to the optimal result. This effect is consistent with other findings in  literature, where has been shown that quantum dynamics in presence of an external perturbation tends to converge towards unitary q-designs (i.e. unitaries that can uniformly cover the Hilbert space) as time increases\cite{banchi2017driven}. This justifies a slower convergence since these types of unitaries are highly expressive and, as previously shown, are much more prone to encountering barren plateaus during optimization\cite{holmes2022connecting, mcclean2018barren}. Interestingly, as reported in Fig.\ref{qsl_fig}c, an initial plateaux in the optimization is present both for very short dynamics and for the ones beyond our estimate $T_{QSL}$. Even though they look similar the former may arise from a lack of controllability (i.e. too few control parameters available), while the latter are a direct manifestation of the barren plateaux. Having identified this sweet spot in terms of the length of the dynamics suggests that, to avoid encountering optimization problems with larger systems, it might be beneficial to progressively optimize the dynamics starting from shorter evolutions and initializing the control parameters to achieve idle evolution. We plan to assess the effect of the initialization and adaptive optimization of the dynamics in a future work.
We refer the reader to appendix\,B where similar results are shown for the LiH molecule.

Finally, to get an additional insight of this multiple interplay among controllability and time-energy uncertainty relationships, we report in Fig.\ref{controllability_qsl} for the H$_4$ molecule the solution of the control problem varying (i) the length of the dynamics (Fig.\ref{controllability_qsl}a-b) and (ii) the number of controllable steps keeping the duration fixed at T=0.01 a.u. (Fig.\ref{controllability_qsl}c-d). Again, we can notice that the amount of energy that the perturbation inputs into the system decreases as the evolution length increases as already shown in Fig.\ref{qsl_fig}. Nevertheless we can notice that moving from T=0.01 to T=0.05 a.u. there is an abrupt decrease as compared to all the values reported both in Fig.\ref{qsl_fig} and Fig.\ref{controllability_qsl}. This motivated us to study the effect of increasing the number of controllable steps at this shorter duration length of the dynamics. As we can see in Fig.\ref{controllability_qsl}c increasing the number of controllable steps immediately leads the control problem to find a state within chemical accuracy. Moreover, the mean energy input into the system decreases of almost two order of magnitudes reaching values more in line with the trends observed for the other systems.
From this calculation arises a picture where the quantity $\langle \frac{||H(t)||}{||H_{mol}||} \rangle$ allows to diagnose possible controllability issues in the definition of the optimization problem. Indeed, its unexpected increase indicates that the optimizer is striving to reach the optimal solution, still attainable within the given time frame. In doing so, it injects more energy into the system but lacks the flexibility to effectively address the control problem. By relaxing this constraint and adding more controllable steps, both the excess injected energy and convergence issues vanish.

\subsection{Computational cost analysis}\label{cost-section}

In this section we provide an empirical estimate of the algorithmic scaling of the proposed method. Particularly, as mentioned in Sec.\ref{classical_optimization} we can approximate the cost (or runtime of the algorithm) as $C=\mathcal{O}(\frac{K M G}{\epsilon^2})$. We recall that K is the number of iterations needed to achieve 1 mHa of error, M is the number of circuits per iteration and G is the gate count of the circuit.

\begin{figure*}[htbp!]

\centering
    \includegraphics[width = \textwidth]{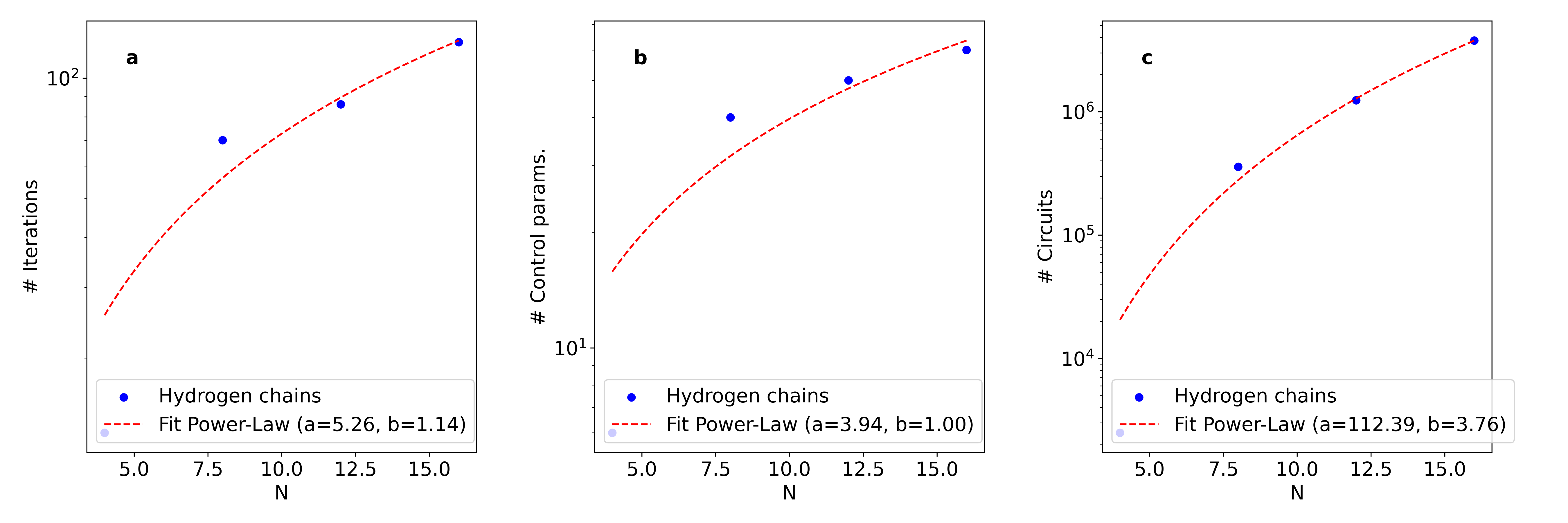}

    \caption{Empirical scaling of the optimal control algorithm. a) Number of iterations needed to reach chemical accuracy as a function of the number of spin-orbitals (N). b) Number of control parameters as a function of the number of spin-orbitals. c) Number of circuits evaluated during the all control problem solution. Power-law fitting function: $y = a N^b$.}
    \label{cost_plot}
\end{figure*}

In Fig.\,\ref{cost_plot}, we provide an estimate of $C$ as a function of the spin-orbitals N studying the hydrogen chain series from H$_2$ to H$_8$ at $r=1\,\AA$. As we can see Fig.\ref{cost_plot}a reports an almost linear scaling of the number of iterations as a function of the spin-orbitals and Fig.\ref{cost_plot}b a linear scaling concerning the number of control parameters. In order to estimate the overall number of circuits executed during the optimization we considered the relation:

\begin{equation}\label{number_of_circuits}
    \# Circuits \approx \mathcal{O}(\# Iterations \, \, \text{x} \, \,  \# params \, \,  \text{x} \, \,  m)
\end{equation} 
Where $m$ is the number of independent measurements needed to compute the 2-RDM of the molecular Hamiltonian according to the shadow procedure developed in Ref.\cite{babbush2023quantum}:

\begin{equation}\label{shadow}
    m = \mathcal{O}(\frac{{4 \eta^2}}{\epsilon^2})
\end{equation}
Where $\eta$ is the number of electrons in the system.

It is important to notice that the estimate of $m$ could be further improved adopting system-dependent methods to reduce the overall number of measurements based on Pauli strings partitioning\cite{martinez2023assessment, choi2023fluid} in combination with matrix completion techniques\cite{peng2023fermionic}.

Plugging Eq.\,\ref{shadow} into Eq.\,\ref{number_of_circuits} we got an overall scaling of $\# Circuits \approx \mathcal{O}(N^4)$. Finally, the total runtime reads:

\begin{equation}
    C = \mathcal{O}(\frac{G N^4}{\epsilon^2})
\end{equation}
In the last equation we decided to keep the runtime of the quantum simulation routine unexpressed as the designated platform for the execution of this algorithm is an analog simulator for the molecular Hamiltonian. Currently, only prototypes of this simulator have been developed\cite{arguello2019analogue, arguello2020quantum}. Nevertheless, if we were to consider implementing the algorithm on digital quantum computers, it would be reasonable (actually conservative) to assume a linear increase in computational overhead as the number of spin orbitals grows\cite{low2023complexity, babbush2019quantum}. 

To conclude this section we would like to comment on our findings. First of all, we are aware that the results of Fig.\,\ref{cost_plot} give rise to a crude estimate for at least two reasons: (i) the modest range of the active space explored and (ii) an additional uncertainty due to the random guess initialization. Moreover, in Sec.\,\ref{oc-energies} we showed that depending on the degree of correlation the number of iterations needed may vary; to this extent, expanding this benchmark to other systems will surely increase its reliability. 

Nevertheless, these results already represent a good starting point to understand if this method is worth further refinements. Given that very promising methods, such as adaptive strategies\cite{grimsley2019adaptive}, have $\mathcal{O}(N^8)$ scaling if implemented naively and can achieve $\mathcal{O}(N^5)$ scaling only if clever strategies for evaluating gradients are adopted\cite{anastasiou2023really}, we think that the method proposed in this paper can be of interest regardless of the future development of an analog simulator. To this extent incremental optimization strategies, i.e. optimizing the evolution step-by-step, could provide significant speedups and provide tighter bounds on the overall scaling. 
We plan to explore these aspects in a future work.

\section{Conclusions}

We have proposed an optimal control approach utilizing a quantum device to steer the evolution of a quantum system towards its ground state. Demonstrating its application, we have computed ground state energies for molecular systems up to 16 spin-orbitals (qubits). Our results indicate the potential to discover pathways reaching states within the chemical accuracy energy threshold more rapidly than the adiabatic path, edging closer to the quantum speed limit. Moreover, this study underscores the intertwined nature of controllability (defined as the minimum number of controls needed to achieve desired precision), duration of the dynamics and convergence of the optimization protocol.

Additionally, we have offered an empirical estimate of the computational cost conducting various calculations on hydrogen chains of different lengths. Our findings reveal that the overall algorithm runtime execution scales as $\mathcal{O}(N^5)$, aligning with results from adaptive ansatze. Future avenues of this work aim to mitigate the computational scaling by refining the procedure via a step-by-step adaptive optimization of the quantum dynamics.

To conclude, we highlight how this method could serve to speed up phase estimation-like algorithms. Particularly the optimally controlled dynamics could impact the quantum phase estimation procedure in two ways: (i) on one hand as initial state preparation routine; on the other hand (ii), we can imagine of replacing the evolution of the circuit within the QPE to directly sample a correlation function (from a simple initial state such as HF) corresponding to an initial state that includes contributions from multiple electronic configurations. We plan to explore this latter aspect in future works to understand whether this option could bring fault-tolerant algorithms' implementation one step closer.

\begin{acknowledgments}
D.C.  is grateful to MIUR ”Dipartimenti di Eccellenza” under the project  Nanochemistry  for  energy  and Health (NExuS) for funding the PhD grant. All the authors acknowledge the CINECA award under the ISCRA C project QuaSiMC (HP10CVSJZR), for the availability of high performance computing resources and support.
\end{acknowledgments}

\section*{Data Availability Statement}

The code that support the findings of this study are openly available at \cite{oc_code}.

\appendix

\section{A simpler example on H$_2$}\label{STA}

\begin{figure*}[htbp!]

\begin{tikzpicture}[node distance=cm,
    every node/.style={fill=white, font=\sffamily}]

\node (figure) at (0,0) {\centering
    \includegraphics[width = \textwidth]{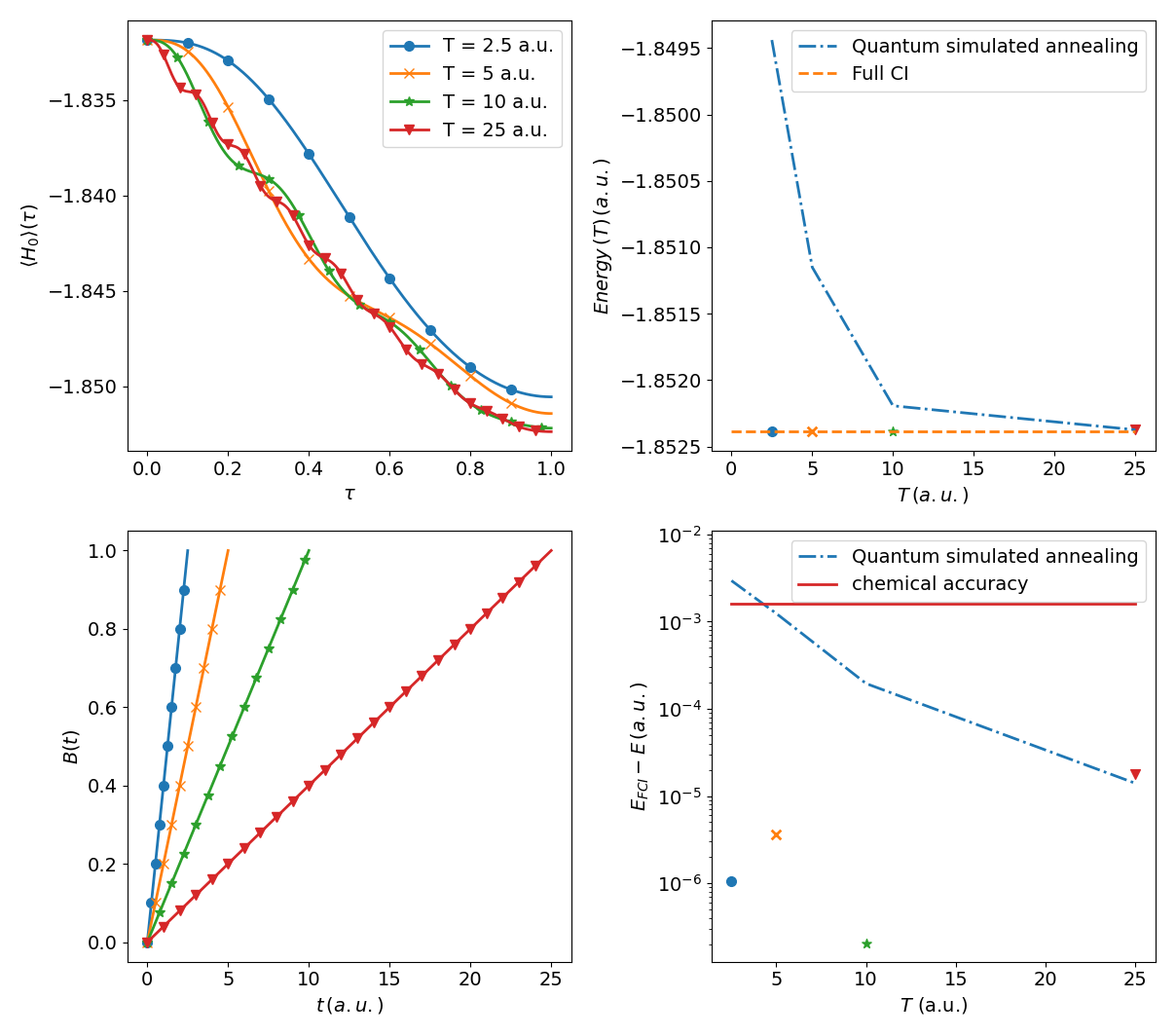}};

\node (a) at (-7.0, 7.0) {a)};

\node (b) at (0.5, 7.0) {b)};

\node (c) at (-7.0, 0) {c)};

\node (d) at (0.5, 0) {d)};

\end{tikzpicture}

    \caption{Comparison between quantum simulated cooling and quantum simulated annealing results for the $H_2$ molecule at equilibrium geometry. a) Expectation value of the driving Hamiltonian (Eq. \ref{annealing_hamiltonian}) as a function of the dimensionless instantaneous time, $\tau = \frac{t}{T}$. Results are reported for different values of annealing time $T$: $T = 2.5$ a.u. (blue dots), $T = 5$ a.u. (orange crosses), $T = 10$ a.u. (green stars) and $T = 25$ a.u. (red downward triangles). b) Electronic energies after quantum simulated annealing (dashed-dotted blue line) and quantum simulated cooling (scatter plot) as a function of the evolution time. Scatter plot symbols refer to the same $T$ values of panel \textit{a}; orange dashed line is the reference FCI energy. c) Linear schedules for the quantum simulated annealing. d) Absolute energy difference from FCI for the quantum simulated cooling (scattered) and quantum simulated annealing (dashed-dotted blue line). The red solid line poses a threshold for chemical accuracy at 0.0016 Ha (i.e., 1 kcal/mol).}
    \label{STA_analysis}
\end{figure*}

The aim of this section is to provide a simpler example of the general protocol, for the case of the hydrogen molecule $H_2$, focusing on the comparison of the results with a quantum simulated annealing protocol for different values of annealing time $T$. We show that the optimal solution of our procedure represents a shortcut to adiabaticity\cite{guery2019shortcuts, martinis2014fast}, i.e. an alternative fast route that allows to obtain the same final state given by a slow, adiabatic evolution.

As to test the optimal control framework in different settings we modified both the choice of the perturbation and the optimization routine (here we used differential evolution as implemented in Scipy\cite{virtanen2020scipy}). Particularly, concerning the control operators here we only needed a time-dependent modulation of the electron-nuclei interaction. Thus, the perturbation operator $V_{\textbf{a}}(t)$ reads:

\begin{equation}
    \hat{V}_{\textbf{a}}(t) =  \sum_{p,q} \tilde{h}^{\textbf{a}}_{pq} (t) a^{\dagger}_{p}a_q
\end{equation}

with $\tilde{h}^{\textbf{a}}_{pq}(t)$ given by:

\begin{equation}
    \tilde{h}^{\textbf{a}}_{pq} (t) = f_{\textbf{a}}(t)\int\phi^*_p(\textbf{x})\sum_i \frac{Z_i}{r_i}\phi_q(\textbf{x})d\textbf{x}
\end{equation}
with $f_{\textbf{a}}$ defined locally in time such that each different value of the function at different time steps of the propagation is a unique control parameter.

The theoretical foundation of quantum simulated annealing is the adiabatic theorem \cite{albash2018adiabatic}. It states that the evolution of a quantum state, being in the ground state of an initial Hamiltonian $\hat{H}_0$, will occur transitionless (i.e. adiabatically, without excitations) under a time dependent Hamiltonian $\hat{H}(t)$ if the variation rate of the Hamiltonian is small enough. As a consequence, a slow evolution under a perturbation that modifies the Hamiltonian until it becomes the one of interest allows to compute its ground state. Here we chose the initial state to be the solution of a classical Hartree-Fock calculation: the initial Hamiltonian is the Hartree-Fock Hamiltonian $\hat{H}_{HF}$. The adiabatic evolution is meant to reach the exact ground state for the complete molecular Hamiltonian $\hat{H}_{mol}$, therefore the overall evolution takes place under the Hamiltonian:

\begin{equation}
\label{annealing_hamiltonian}
    \hat{H}(t) = A(t)\hat{H}_{HF} + B(t)\hat{H}_{mol}
\end{equation}
The functions $A(t)$ and $B(t)$ define the annealing schedule, i.e. the switching factors between the two Hamiltonians. They must be defined such that $\hat{H}(0) = \hat{H}_{HF}$ and $\hat{H}(T) = \hat{H}_{mol}$.

The choice of the annealing schedule influences the perfomance of the quantum algorithm\cite{roland2002quantum, albash2015decoherence, chen2020and}. In particular, different strategies have been adopted such as optimal control protocols devised to shorten the annealing time or to shape the annealing schedule profile enhancing the success probability \cite{brady2021optimal, chen2020optimizing}. Bearing this in mind, here we have considered a linear schedule (Eq. \ref{annealing_schedule}) as it is still used as typical benchmark in the field \cite{callison2021}. 


\begin{equation}
\label{annealing_schedule}
\begin{split}
 A(t) = 1 -&  \frac{t}{T} \\ B(t) = \frac{t}{T} 
\end{split}
\end{equation}

In Fig. \ref{STA_analysis} we show the results of our comparison between quantum simulated cooling and quantum simulated annealing, for this analysis we focused on the $H_2$ molecule at the equilibrium distance. We considered four increasing evolution times ranging from 2.5 a.u. to 25 a.u.. In the first panel (Fig. \ref{STA_analysis}a), the driving Hamiltonian expectation value is plotted as a function of the instantaneous time, $\tau = \frac{t}{T}$. As expected, increasing the annealing time the energy at the final instant decreases as the evolution occurs without transitions to any excited electronic configuration. Moving to panel \ref{STA_analysis}b we compare the results of the adiabatic evolution (dashed-dotted blue line) with the exact Full CI energy (dashed orange line) and the energy outcome of the quantum simulated cooling (scatter plot symbols differ according to the evolution length). We observe that the non-adiabatic evolution provided by the quantum simulated cooling is always performing better than the quantum simulated annealing protocol. The only exception is for the last point at 25 a.u., which, as better highlighted in Fig. \ref{STA_analysis}d, equals the result obtained with the Hamiltonian of Eq. \ref{annealing_hamiltonian}. To explain this behavior we point out that the differential evolution is a non-deterministic algorithm, as such small fluctuations on the final results are expected between different runs.

\section{Additional calculations on the LiH molecule}\label{lih_other_calculations}

\begin{figure*}[htbp!]

\centering
    \includegraphics[width = \textwidth]{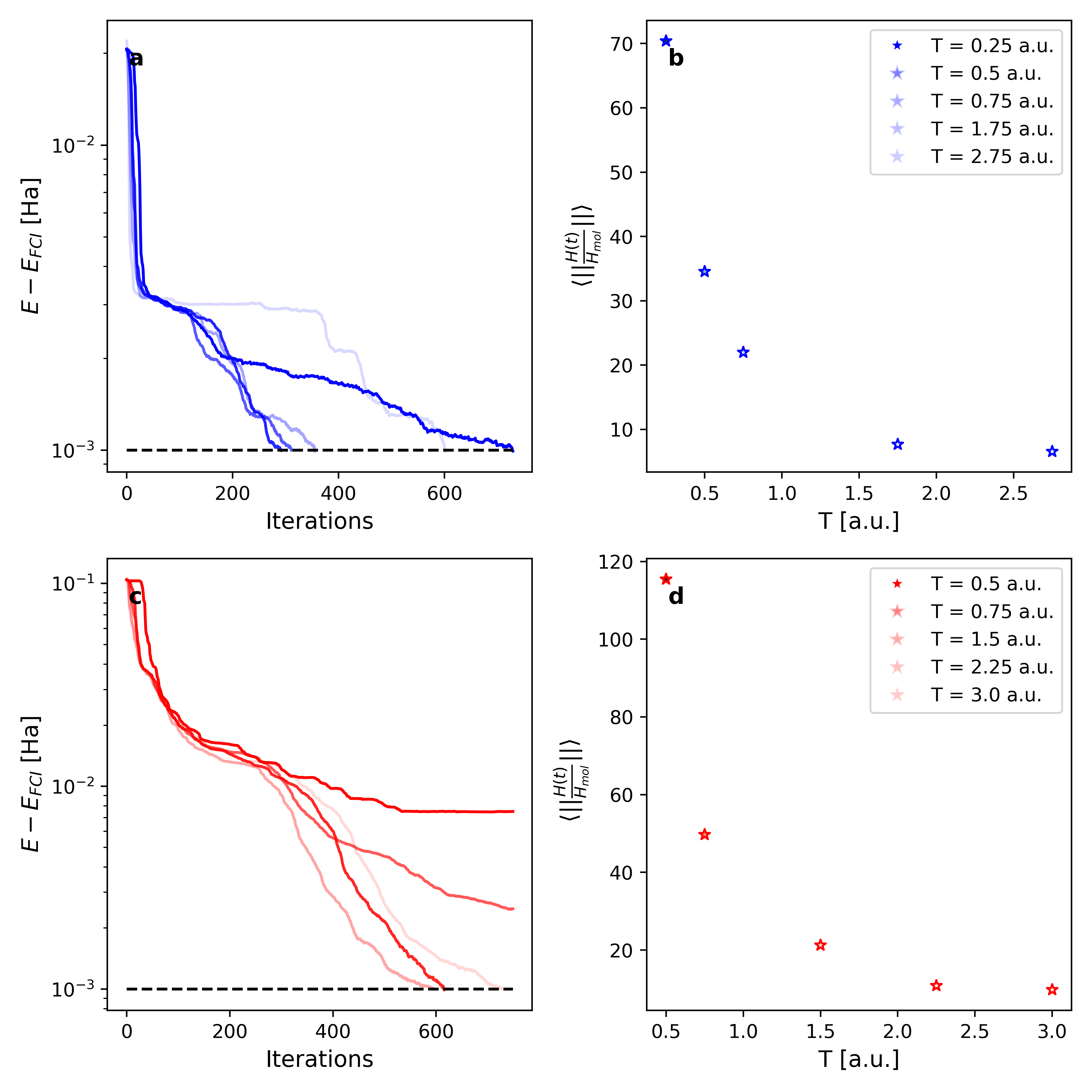}

    \caption{Ground state preparation at the quantum speed limit for the LiH molecule. a, c) Optimal control for various duration of the dynamics. Blues refer to LiH at r=1.6 $\AA$, reds refer to LiH at r=3.2 $\AA$. b,d) Mean driving Hamiltonian norm vs. duration length for the same systems and color codes.}
    \label{qsl_fig_lih}
\end{figure*}

Here we report additional results on the LiH molecule (see Fig.\ref{qsl_fig_lih}) concerning the analysis on the dynamics' duration length as discussed in Sec.\ref{qsl-sec}. These results confirm the considerations drawn in the main text: short evolution lengths imply higher energy injection from the perturbation which results in higher values of $\langle \frac{||H(t)||}{||H_{mol}||} \rangle$. Concerning the convergence behavior of the control problem w.r.t. the length of the evolution at the equilibrium geometry (Fig.\ref{qsl_fig_lih}a) we can clearly see the same pattern previously discussed, i.e. that very short (or very long) evolutions can hamper the optimization to reach a chemically accurate result. On the other hand, this trend is not so evident looking at Fig.\ref{qsl_fig_lih}c (r = 3.2 $\AA$). The dynamics' length and convergence patterns appear less interdependent in this system. Further studies collecting statistics on various randomly initialized optimizations could resolve this inconsistency.

\nocite{*}
\bibliography{aipsamp}

\end{document}